\documentclass[sigconf]{acmart}

\usepackage{enumitem}
\usepackage{tikz}
\usepackage{amsmath}
\usepackage{booktabs}
\usepackage{hhline}
\usepackage{multirow}
\usepackage{pdflscape}
\usepackage{makecell}
\usepackage{array}
\usepackage{placeins}
\usepackage{subcaption}
\usepackage{listings}
\usepackage[table]{xcolor}
\usepackage{tcolorbox}

\AtBeginDocument{%
  }

\setcopyright{acmlicensed}
\copyrightyear{2026}
\acmYear{2026}
\acmDOI{XXXXXXX.XXXXXXX}
\acmConference[ASE]{Make sure to enter the correct
  conference title from your rights confirmation email}{October 12--16,
  2026}{Munich, Germany}
\acmISBN{978-1-4503-XXXX-X/2018/06}
  
\acmJournal{JACM}
\acmVolume{37}
\acmNumber{4}
\acmArticle{111}
\acmMonth{8}

\definecolor{hm0}{rgb}{0.56, 0.87, 0.56}
\definecolor{hm1}{rgb}{0.76, 0.90, 0.54}
\definecolor{hm2}{rgb}{0.99, 0.88, 0.51}
\definecolor{hm3}{rgb}{0.99, 0.70, 0.42}
\definecolor{hm4}{rgb}{0.89, 0.26, 0.20}
\newcommand{\hmc}[1]{%
  \ifnum#1=0\relax\cellcolor{hm0}%
  \else\ifnum#1<4\relax\cellcolor{hm1}%
  \else\ifnum#1<6\relax\cellcolor{hm2}%
  \else\ifnum#1<10\relax\cellcolor{hm3}%
  \else\cellcolor{hm4}%
  \fi\fi\fi\fi#1}

\definecolor{codeblue}{rgb}{0.12, 0.47, 0.71}
\definecolor{codegreen}{rgb}{0, 0.6, 0}
\definecolor{codegray}{rgb}{0.5, 0.5, 0.5}
\definecolor{codepurple}{rgb}{0.58, 0, 0.82}
\definecolor{codered}{rgb}{0.70, 0.09, 0.17}
\definecolor{backcolour}{rgb}{0.95, 0.95, 0.92}

\lstdefinestyle{codesnippet}{
  basicstyle=\ttfamily\fontsize{4.5}{5.5}\selectfont,
  columns=fullflexible,
  breaklines=true,
  breakatwhitespace=true,
  keepspaces=true,
  showstringspaces=false,
  commentstyle=\itshape\color{gray!70},
  keywordstyle=\color{black},
}

\lstdefinestyle{mystyle}{
  backgroundcolor=\color{backcolour},
  basicstyle=\ttfamily\fontsize{4.5}{5.5}\selectfont,
  breaklines=true,
  keepspaces=true,
  showstringspaces=false,
  captionpos=t,
  numbers=left,
  numberstyle=\fontsize{4}{5}\selectfont\color{codegray},
  numbersep=6pt,
  frame=single,
  rulecolor=\color{codegray},
  commentstyle=\color{codegreen}\itshape,
  keywordstyle=\color{codepurple}\bfseries,
  stringstyle=\color{codered},
  columns=fullflexible,
}

\definecolor{customblue}{HTML}{006ca6}
\definecolor{customgreen}{HTML}{009264}
\definecolor{custombrown}{HTML}{ff3d00}
\AtEndPreamble{
 \usepackage{hyperref}
 \hypersetup{
  colorlinks = true,
  linkcolor = customblue,
  anchorcolor = customblue,
  citecolor = customgreen,
  filecolor = customblue,
  urlcolor = customblue
 }
}

\newcommand{\find}[1]{
\begin{tcolorbox}[leftrule=1mm,toprule=0mm,bottomrule=0mm,left=1pt,right=2pt,top=2pt,bottom=2pt]
\em #1
\end{tcolorbox}
}

\authorsaddresses{} 
\renewcommand\footnotetextcopyrightpermission[1]{}

\begin{document}

\title{\textit{Many a Little Makes a Mickle}: A Code-Centric Empirical Study of Data Minimization Principle in Android App Development}

\newcommand{\tool}{ourTool}


\author{Dianshu Liao}
\affiliation{
  \institution{Australian National University}
  \country{Australia}
}

\author{Shidong Pan}
\authornote{Shidong Pan completed this work while he was a visiting research scientist at CSIRO's Data61.}
\affiliation{
  \institution{CSIRO}
  \country{Australia}
}

\author{Zhenchang Xing}
\affiliation{
  \institution{CSIRO \& Australian National University}
  \country{Australia}
}

\author{Xiaoyu Sun}
\authornote{Corresponding author.}
\affiliation{
  \institution{Australian National University}
  \country{Australia}
}

\lstdefinelanguage{JavaScript}{
  keywords={typeof, new, true, false, catch, function, return, null, catch, switch, var, if, in, while, do, else, case, break},
  keywordstyle=\color{codepurple}\bfseries,
  ndkeywords={class, export, boolean, throw, implements, import, this},
  ndkeywordstyle=\color{codeblue}\bfseries,
  identifierstyle=\color{black},
  sensitive=false,
  comment=[l]{//},
  morecomment=[s]{/*}{*/},
  morestring=[b]',
  morestring=[b]"
}

\begin{abstract}
Modern mobile applications consume large amounts of data to function, raising significant privacy concerns and regulatory challenges.
While prior work has primarily focused on detecting compliance gaps through policy analysis, there remains a lack of actionable guidance for developers to implement privacy principles at the code level. 
In this paper, we focus on \emph{data minimization} as a developer-operationalizable principle and investigate its realization in Android applications.
We conduct a formative study on 1,114 open-source Android apps to identify ten recurring data minimization scenarios across five data-handling stages. 
Building on this, we perform a large-scale analysis of 9,875 real-world APKs and distill 31 actionable coding guidelines to support privacy-compliant development.
We further examine LLM-based code generation in Android development and find that state-of-the-art models consistently reproduce data minimization–risky practices, indicating that they inherit and amplify patterns from real-world code. 
Encouragingly, incorporating our guidelines eliminates these issues across all evaluated models.
Our work advocates a shift toward responding to privacy regulatory requirements at their code-level root causes, enabling better compliance in both human and AI-assisted programming.

\end{abstract}

\maketitle
\fancyhead{}   
\section{Introduction}

Mobile applications have deeply integrated into nearly every aspect of daily life, empowering services such as transportation, personal finance, and social networking~\cite{Liao2025ACS}.
To deliver these services effectively, applications need to consume large amounts of data, such as location, names, email addresses, and account details. 
This extensive data handling has raised growing concerns that the collected data might be misused, exposing users to risks such as privacy leaks~\cite{Gibler2012AndroidLeaksAD,WangGUILeak,sun2021characterizing,pan2024trap}, identity theft~\cite{Fedynyshyn2025DataPA}, and persistent harassment~\cite{Dhondt2024SwipeLF}.
These concerns are further amplified by rapid AI-assisted development practices (e.g., vibe coding), where developers may prioritize functionality over careful privacy engineering~\cite{Klemmer2024UsingAA,Haque2025SOKEH,jin2026understanding}.

In response, governments and regulatory bodies have introduced privacy frameworks such as the General Data Protection Regulation (GDPR)~\cite{GDPR} to regulate the collection and processing of personal data. 
Existing research has extensively studied privacy compliance in mobile applications by comparing observed data practices with privacy disclosures, such as privacy policies or labels~\cite{Fan2020AnEE,Zimmeck2016AutomatedAO,Slavin2016TowardAF,tao2025privacy, si2024solution, pan2024new, pan2023toward}. 
Through such analysis, researchers identify noncompliance in large-scale, for example, cases where an application collects sensitive data (e.g., location information) without adequately disclosing such practices in its privacy policy~\cite{yu2016can, Zimmeck2016AutomatedAO, yu2018ppchecker, andow2020actions, tao2025longitudinal}.
Although this line of research has been effective in evaluating privacy compliance and advancing the state-of-the-art, the identified instances of noncompliance or misalignment are often ad-hoc and offer limited practical guidance for developers or for informing future mobile software engineering practices.
In practice, developers implement numerous concrete coding decisions at the source-code level to realize application functionality~\cite{sommerville2011software}.
However, high-level privacy regulations do not provide actionable guidance for writing code that complies with principles such as data minimization~\cite{Ferreira2023RuleKeeperGP,Klein2023GeneralDP}.
During development, it is therefore difficult to assess an application’s overall privacy compliance, and it is unrealistic to expect developers to directly operationalize abstract regulatory requirements in day-to-day implementation.

Notably, data minimization is a privacy principle that can be directly operationalized at the implementation level~\cite{Senarath2019ADM,Senarath2018UnderstandingSD}. 
This feasibility stems from the fact that decisions about what data to create, collect, transmit, store, and retain are concretely realized in code, and are therefore under the direct control of developers during implementation.
Intuitively, if individual code segments adhere to data minimization practices, the overall application is less likely to exhibit privacy noncompliance.
Nevertheless, despite this apparent tractability, developers may fail to correctly operationalize data minimization due to a lack of fine-grained guidance on how to implement the principle in practice.
Thus, this paper focuses on data minimization as a proactive developer-oriented aspect of privacy engineering at code-level in Android application development.

Specifically, through a formative study of 1,114 open-source Android applications, we identify ten recurring data minimization scenarios across five data-handling stages.
Building on these findings, we conduct a heuristic-based large-scale analysis of 9,875 Android APKs from AndroZoo~\cite{liu2020androzooopen}, focusing on apps with at least 10 million downloads.
We find that 7,548 APKs (76.44\%) exhibit at least one data minimization–related signal, such as unnecessary sensitive permission requests, APIs that may cause over-collection, data flows from sensitive sources to network sinks, and improper data handling configurations.
Based on these results, we characterize recurring data handling patterns and distill 31 actionable coding guidelines to support data minimization–compliant development.

As LLM-based coding becomes a dominant paradigm, we evaluate whether these approaches generate data minimization–risky code.
We assess LLM-based code generation across 30 Android scenarios covering all five data handling stages.
Our results show that advanced LLMs (Gemini-2.5-Pro~\cite{gemini25pro}, GPT-5.2~\cite{GPT52}, Claude-4.5-Sonnet~\cite{claude45sonnet}) and the coding tool Cursor~\cite{Cursor} produce 46–59 risky instances, suggesting that models learn and propagate non-compliant practices from real-world code.
Encouragingly, incorporating our guidelines reduces these issues to zero across all approaches. 
These findings indicate that LLMs amplify existing coding practices but can be steered toward privacy-compliant implementations through domain-specific guidance.


Overall, this paper makes the following major contributions:
\begin{itemize}[leftmargin=*]
    \item We conduct a formative study on 1,114 open-source Android applications and identify ten recurring code-level data minimization scenarios, which are then operationalized and analyzed at scale on 9,875 real-world APKs to characterize prevalent data handling coding patterns.
    \item Based on large-scale empirical characterization, we distill 31 actionable coding guidelines that help developers implement data minimization in practice.
    \item We show that modern LLM-based coding approaches inherit and reproduce data minimization-risky practices from real-world code, and demonstrate that incorporating our guidelines effectively guides models toward privacy-compliant implementations.
\end{itemize}

\section{Background and Motivation}
\paragraph{Data Minimization in Privacy Regulations}
In response to privacy risks from excessive data handling in mobile applications, governments and regulatory bodies have introduced data protection regulations, including the European Union’s GDPR~\cite{GDPR}, California’s CCPA~\cite{CCPA}, and China’s Personal Information Protection Law (PIPL)~\cite{PIPL}. 
Despite differences in legal formulations, these regulations converge on a common requirement on data minimization: 
collected data should be limited to what is necessary in relation to the purpose~\cite{Klein2023GeneralDP,Ferreira2023RuleKeeperGP,tao2025privacy}.
The less data is generated, collected, transmitted, stored, and retained, the lower the risk of misuse (whether malicious or unintentional) and the consequences of data breaches.




\paragraph{Android Privacy Control Framework}

Android provides a privacy control infrastructure to regulate applications’ access to privacy-sensitive data, such as location, device identifiers, and contacts.
This infrastructure is based on a permission framework that determines whether an application can access protected data~\cite{pan2025first, AndroidPermissionsOverview, Felt2011AndroidPD}.
It primarily specifies what data can be accessed by developers, while leaving them free to determine how the data is subsequently handled within the application.
Starting from Android 6.0, the introduction of runtime permissions further strengthened user control by requiring applications to request sensitive permissions during execution rather than at installation time~\cite{AndroidRuntimePermissions}.
Although this mechanism significantly improves transparency and user awareness, it does not enforce constraints on downstream data-handling practices, thereby leaving substantial privacy responsibility to developers.

\paragraph{Code-centric View of Data Lifecycle in Android Applications.}
Prior work in privacy engineering and data protection regulations has characterized personal data handling as a lifecycle involving stages such as collection, processing, storage, and sharing~\cite{Grses2011EngineeringPB,Solove2006ATO,Canedo2022PrivacyRE,wang2025big}.
In mobile applications, these stages are further elaborated by platform-specific mechanisms such as permission-based access control and runtime data flows~\cite{Felt2012AndroidPU,Zimmeck2016AutomatedAO,Reardon201950WT}.
Based on these foundations and our observations of Android application code, we operationalize a \emph{five-stage code-centric view} of data handling.

The process begins with \textit{permission control}, indicating whether an application is granted the capability to access certain data through the permission. 
Once access is granted, applications proceed to \textit{data acquisition}, which concerns the scope of data collection, i.e., what data is acquired, how much data is collected, and at what level of granularity for subsequent application logic. 
The acquired data may then undergo \textit{data transmission} to implement application functionality for processing, either within the application code itself or via third-party libraries. 
Following processing, data may enter the \textit{data storage} stage, where it is persisted for future use in forms such as databases, local files, or cloud storage services. 
Finally, in the \textit{data retention} stage, stored data may be retained under time-bounded policies (e.g., retention with expiry), along with mechanisms for deletion or cleanup once the data is no longer needed.

Despite the existence of regulatory requirements and platform-level protections, there remains a lack of concrete, implementation-level guidance for developers on how to realize data minimization principles in practice.
Existing coding standards and style guides (e.g.,~\cite{GoogleCodeStyle}) provide extensive recommendations for code quality and maintainability, yet analogous guidance for privacy-aware implementation is largely absent. 
In this work, we aim to bridge this gap by systematically identifying actionable coding practices that operationalize data minimization in Android application development.

\section{Formative Study}
\label{sec_formative_study}

We conduct a formative study on open-source Android applications by qualitatively examining data-handling snippets in source code to elicit scenarios about data minimization (DM).

\subsection{Qualitative Observation Procedure}
Following prior empirical software engineering studies, we adopt an iterative qualitative coding process inspired by grounded theory~\cite{Runeson2009GuidelinesFC, Stol2016GroundedTI} to derive a taxonomy of DM-relevant coding practices. 
We select F-Droid~\cite{FDroid} as the data source.
F-Droid is a repository of open-source Android applications across 55 categories, and it has been widely used in mobile software engineering research~\cite{Liao2025ACS,SunTR2021}.
To ensure representativeness, we randomly sample 30 applications per category, and for categories with less than 30 applications, we collect all available applications.
In total, we obtain 1,114 Android applications.
For each app, we extract the source code of its latest version and conduct the empirical observation as follows:

\textbf{Step 1: Locate data-related code segments.}
For each sampled application, we locate and identify code segments and configuration artifacts related to data handling.
Specifically, we inspect Android permission declarations, sensitive API usages, local data persistence operations (e.g., databases, files, and shared preferences), logging behaviors, and interactions with third-party libraries.

\textbf{Step 2: Manual inspection and open coding.}
Two authors independently inspect the identified instances and perform open coding to characterize DM-relevant decision points and implementation patterns.
Example instances include retrieving data from local databases, accessing shared preferences, obtaining user input, and persisting data to files or databases (Table~\ref{tab_dm_stage_scenarios}).
Each instance is assigned an initial code describing its data-handling behavior.

\textbf{Step 3: Pattern clustering and codebook refinement.}
Two authors iteratively reconcile disagreements and cluster similar initial codes into higher-level patterns organized around the five data-handling stages. 
Throughout this process, a codebook is maintained and refined to define each stage-aligned pattern and to distinguish between \emph{DM-aligned} and \emph{DM-risky} instances.

\textbf{Step 4: Taxonomy construction.}
We iterate steps 2 and 3 until reaching saturation, i.e., when additional applications no longer reveal substantively new DM-relevant patterns.
Finally, we synthesize a taxonomy by refining the observations into ten recurring DM-critical coding scenarios across five stages.

\begin{table*}[]
    \centering
    \scriptsize
    \setlength{\tabcolsep}{4pt}
    \renewcommand{\arraystretch}{1.15}
    \caption{A stage-structured taxonomy of DM scenarios with representative \emph{DM-aligned} and \emph{DM-risky} coding practices.}
    \label{tab_dm_stage_scenarios}
    \vspace{-4mm}
    \begin{tabular}{@{}>{\centering\arraybackslash}m{0.03\textwidth} | p{0.07\textwidth} | p{0.4\textwidth} | p{0.4\textwidth}@{}}
    \hline
    \textbf{Stage} & \textbf{DM Scenario} & \textbf{\makecell[l]{DM-aligned Practices}} & \textbf{\makecell[l]{DM-risky Practices}} \\
    \hline
    \multirow{2}{*}[-10.0ex]{\rotatebox[origin=c]{90}{\makecell[c]{S1- Permission Control}}} & 1.1 Permission Declaration Minimization &
    Declare only permissions exercised by the code.\par\vspace{-2pt}
    \begin{minipage}[t]{\linewidth}\vspace{0pt}
    \begin{lstlisting}[style=mystyle,language=Java,numbers=none,xleftmargin=0pt,framexleftmargin=0pt,framesep=2pt,aboveskip=0pt,belowskip=0pt]
// Querying CalendarContract.Events requires READ_CALENDAR.
Cursor cursor = contentResolver.query(CalendarContract.Events.CONTENT_URI, ...);

<!-- AndroidManifest.xml -->
<uses-permission android:name="android.permission.READ_CALENDAR" />\end{lstlisting}\vspace{1pt}
    \end{minipage} &
    Declare unused permissions.\par\vspace{-2pt}
    \begin{minipage}[t]{\linewidth}\vspace{0pt}
    \begin{lstlisting}[style=mystyle,language=Java,numbers=none,xleftmargin=0pt,framexleftmargin=0pt,framesep=2pt,aboveskip=0pt,belowskip=0pt]
// Only queries calendar events. READ_CONTACTS is unnecessary.
Cursor cursor = contentResolver.query(CalendarContract.Events.CONTENT_URI, ...);
<!-- AndroidManifest.xml -->
<uses-permission android:name="android.permission.READ_CALENDAR" />
<uses-permission android:name="android.permission.READ_CONTACTS" />\end{lstlisting}\vspace{1pt}
    \end{minipage} \\
    
    \cline{2-4}
    & 1.2 Permission Request Minimization &
    Request permissions at point of use (feature-triggered).\par\vspace{-2pt}
    \begin{minipage}[t]{\linewidth}\vspace{0pt}
    \begin{lstlisting}[style=mystyle,language=Java,numbers=none,xleftmargin=0pt,framexleftmargin=0pt,framesep=2pt,aboveskip=0pt,belowskip=0pt]
public class MainActivity extends AppCompatActivity {
  @Override
  protected void onCreate(Bundle savedInstanceState) {
    super.onCreate(savedInstanceState);
    setContentView(R.layout.activity_main); }
  private void showNavigation() {
    if (ContextCompat.checkSelfPermission(...) {...}
    // startNavigation();
  }}\end{lstlisting}\vspace{1pt}
    \end{minipage} &
    Request permissions preemptively (e.g., at app launch).\par\vspace{-2pt}
    \begin{minipage}[t]{\linewidth}\vspace{0pt}
    \begin{lstlisting}[style=mystyle,language=Java,numbers=none,xleftmargin=0pt,framexleftmargin=0pt,framesep=2pt,aboveskip=0pt,belowskip=0pt]
public class MainActivity extends AppCompatActivity {
  @Override
  protected void onCreate(Bundle savedInstanceState) {
    ...
    // Requests location permission when the app starts.
    if (ContextCompat.checkSelfPermission(this, Manifest.permission.ACCESS_FINE_LOCATION) != PackageManager.PERMISSION_GRANTED) {
      ActivityCompat.requestPermissions(this, new String[]{Manifest.permission.ACCESS_FINE_LOCATION}, 1);}}}\end{lstlisting}\vspace{1pt}
\end{minipage} \\
    \hline
    \multirow{1}{*}[-0.8ex]{\rotatebox[origin=c]{90}{\makecell[c]{\tiny{S2 - Acquisition}}}} & 2.1 Data  \/ \/ Acquisition Minimization &
    Retrieve only required fields.\par\vspace{-2pt}
    \begin{minipage}[t]{\linewidth}\vspace{0pt}
    \begin{lstlisting}[style=mystyle,language=Java,numbers=none,xleftmargin=0pt,framexleftmargin=0pt,framesep=2pt,aboveskip=0pt,belowskip=0pt,emph={projection},emphstyle=\color{codegreen}]
String[] projection = new String[] {ContactsContract.Contacts._ID,ContactsContract.Contacts.DISPLAY_NAME}; // Specify the required fields.
Cursor c = getContentResolver().query(ContactsContract.Contacts.CONTENT_URI, projection, null, null, null);\end{lstlisting}\vspace{1pt}
    \end{minipage} &
    Retrieve all available fields.\par\vspace{-2pt}
    \begin{minipage}[t]{\linewidth}\vspace{0pt}
    \begin{lstlisting}[style=mystyle,language=Java,numbers=none,xleftmargin=0pt,framexleftmargin=0pt,framesep=2pt,aboveskip=0pt,belowskip=0pt]
Cursor c = getContentResolver().query(ContactsContract.Contacts.CONTENT_URI,
        null, // projection == null returns all columns.
        null, null, null);
    \end{lstlisting}\vspace{1pt}
\end{minipage} \\
    
    \hline
    \multirow{1}{*}[-0.2ex]{\rotatebox[origin=c]{90}{\makecell[c]{\fontsize{4.8}{5.8}\selectfont{S3 - Transmission}}}} 
    & 3.1 Data Transmission Minimization &
    Transmit only the minimum sensitive data necessary to external endpoints.\par\vspace{-2pt}
    \begin{minipage}[t]{\linewidth}\vspace{0pt}
    \begin{lstlisting}[style=mystyle,language=Java,numbers=none,xleftmargin=0pt,framexleftmargin=0pt,framesep=2pt,aboveskip=0pt,belowskip=0pt]
// Strip optional PII before sending to external endpoint.
String url = "https://example.com/article?id=1&page=1&email=alice@example.com";
String cleanUrl = stripPII(url); // "https://example.com/article?id=1&page=1"
serverAPI.saveBookmark(cleanUrl);\end{lstlisting}\vspace{1pt}
\end{minipage} &
    Transmit sensitive data beyond what the functionality requires.\par\vspace{-2pt}
    \begin{minipage}[t]{\linewidth}\vspace{0pt}
    \begin{lstlisting}[style=mystyle,language=Java,numbers=none,xleftmargin=0pt,framexleftmargin=0pt,framesep=2pt,aboveskip=0pt,belowskip=0pt]
// Directly send URL with PII to external endpoint.
String url = "https://example.com/article?id=1&page=1&email=alice@example.com";
serverAPI.saveBookmark(url);
    \end{lstlisting}\vspace{1pt}
    \end{minipage} \\

    \hline
    \multirow{6}{*}[-40ex]{\rotatebox[origin=c]{90}{\makecell[c]{S4 - Storage}}} & 4.1 Data Logging Minimization &
    Disable verbose logs in release builds.\par\vspace{-2pt}
    \begin{minipage}[t]{\linewidth}\vspace{0pt}
    \begin{lstlisting}[style=mystyle,language=Java,numbers=none,xleftmargin=0pt,framexleftmargin=0pt,framesep=2pt,aboveskip=0pt,belowskip=0pt]
if (BuildConfig.DEBUG) {Log.d("Auth", "token=" + token);}\end{lstlisting}\vspace{1pt}
    \end{minipage} &
    Log sensitive data without restriction.\par\vspace{-2pt}
    \begin{minipage}[t]{\linewidth}\vspace{0pt}
    \begin{lstlisting}[style=mystyle,language=Java,numbers=none,xleftmargin=0pt,framexleftmargin=0pt,framesep=2pt,aboveskip=0pt,belowskip=0pt]
Log.d("Auth", "token=" + token);\end{lstlisting}\vspace{1pt}
    \end{minipage} \\
    
    \cline{2-4}
    & 4.2 Data Backup Minimization &
    Restrict backup scope via rules or directly disable backups.\par\vspace{-2pt}
    \begin{minipage}[t]{\linewidth}\vspace{0pt}
    \begin{lstlisting}[style=mystyle,language=XML,numbers=none,xleftmargin=0pt,framexleftmargin=0pt,framesep=2pt,aboveskip=0pt,belowskip=0pt]
<!-- AndroidManifest.xml -->
<application android:allowBackup="true" android:fullBackupContent="@xml/backup_rules" />
<!-- res/xml/backup_rules.xml -->
<full-backup-content>
  <!-- exclude sensitive data -->
  <exclude domain="sharedpref" path="auth_data" />
  <!-- only include non-sensitive data -->
  <include domain="sharedpref" path="settings" />
</full-backup-content>\end{lstlisting}\vspace{1pt}
    \end{minipage} &
    Enable backups without restricting scope.\par\vspace{-2pt}
    \begin{minipage}[t]{\linewidth}\vspace{0pt}
    \begin{lstlisting}[style=mystyle,language=XML,numbers=none,xleftmargin=0pt,framexleftmargin=0pt,framesep=2pt,aboveskip=0pt,belowskip=0pt]
<!-- AndroidManifest.xml -->
<application android:allowBackup="true" />

<!-- No backup_rules.xml configured -->
<!-- No data_extraction_rules.xml configured -->



    \end{lstlisting}\vspace{1pt}
    \end{minipage} \\
    
    \cline{2-4}
    
    \cline{2-4}
    & 4.3 Key Storage Minimization &
    Store keys in Android KeyStore (HSM/TEE-backed, non-exportable).\par\vspace{-2pt}
    \begin{minipage}[t]{\linewidth}\vspace{0pt}
    \begin{lstlisting}[style=mystyle,language=Java,numbers=none,xleftmargin=0pt,framexleftmargin=0pt,framesep=2pt,aboveskip=0pt,belowskip=0pt]
// Keys stored in AndroidKeyStore (harder to extract even if rooted).
KeyGenerator keyGenerator = KeyGenerator.getInstance(
  KeyProperties.KEY_ALGORITHM_AES, "AndroidKeyStore");
KeyGenParameterSpec spec = new KeyGenParameterSpec.Builder(...).build();
keyGenerator.init(spec);
keyGenerator.generateKey();\end{lstlisting}\vspace{1pt}
    \end{minipage} &
    Persist exportable key material in app storage (readable if rooted).\par\vspace{-2pt}
    \begin{minipage}[t]{\linewidth}\vspace{0pt}
    \begin{lstlisting}[style=mystyle,language=Java,numbers=none,xleftmargin=0pt,framexleftmargin=0pt,framesep=2pt,aboveskip=0pt,belowskip=0pt]
// Exportable key bytes persisted in SharedPreferences (readable if rooted).
SecretKey secretKey = KeyGenerator.generateKey();
String encodedKey = Base64.encodeToString(...);
SharedPreferences prefs = context.getSharedPreferences(...);
prefs.edit().putString("encoded_key", encodedKey).apply();
    \end{lstlisting}\vspace{1pt}
    \end{minipage} \\
    
    \cline{2-4}
    & 4.4 Biometric Handling Minimization &
    Use \emph{BiometricPrompt} to receive outcomes only.\par\vspace{-2pt}
    \begin{minipage}[t]{\linewidth}\vspace{0pt}
    \begin{lstlisting}[style=mystyle,language=Java,numbers=none,xleftmargin=0pt,framexleftmargin=0pt,framesep=2pt,aboveskip=0pt,belowskip=0pt]
// Use Android BiometricPrompt API.
BiometricPrompt prompt = new BiometricPrompt(...) {
    @Override
    public void onAuthenticationSucceeded(AuthenticationResult result) {
      // Application only receives authentication result (success).
      unlockStorage();}});
prompt.authenticate(promptInfo);\end{lstlisting}\vspace{1pt}
    \end{minipage} &
    Application directly stores and manages biometric data.\par\vspace{-2pt}
    \begin{minipage}[t]{\linewidth}\vspace{0pt}
    \begin{lstlisting}[style=mystyle,language=Java,numbers=none,xleftmargin=0pt,framexleftmargin=0pt,framesep=2pt,aboveskip=0pt,belowskip=0pt]
// Application directly stores biometric data.
class BiometricManager {
  void storeFingerprint(byte[] fingerprintData) {
    SharedPreferences.edit().putString(
      "fingerprint", Base64.encode(fingerprintData));
  }
}\end{lstlisting}\vspace{1pt}
    \end{minipage} \\
    
    \cline{2-4}
    & 4.5 Data Identifiability Minimization &
    Encrypt stored account data (e.g., credentials and API keys).\par\vspace{-2pt}
    \begin{minipage}[t]{\linewidth}\vspace{0pt}
    \begin{lstlisting}[style=mystyle,language=Java,numbers=none,xleftmargin=0pt,framexleftmargin=0pt,framesep=2pt,aboveskip=0pt,belowskip=0pt]
// EncryptedSharedPreferences (values encrypted at rest).
MasterKey masterKey = new MasterKey.Builder(context).setKeyScheme(MasterKey.KeyScheme.AES256_GCM).build();
SharedPreferences prefs = EncryptedSharedPreferences.create(context, "auth", masterKey, ...);
prefs.edit().putString(...).apply();\end{lstlisting}\vspace{1pt}\vspace{1pt}
    \end{minipage} &
    Store account data in plaintext.\par\vspace{-2pt}
    \begin{minipage}[t]{\linewidth}\vspace{0pt}
    \begin{lstlisting}[style=mystyle,language=Java,numbers=none,xleftmargin=0pt,framexleftmargin=0pt,framesep=2pt,aboveskip=0pt,belowskip=0pt]
// SharedPreferences (plaintext at rest).
SharedPreferences prefs = context.getSharedPreferences("auth", Context.MODE_PRIVATE);
prefs.edit().putString(...).apply();

    
    \end{lstlisting}\vspace{1pt}
    \end{minipage} \\
    \hline
    \multirow{1}{*}[-4ex]{\rotatebox[origin=c]{90}{\makecell[c]{S5 - Retention}}} & 5.1 Retention Time Minimization &
    Apply TTL/expiry and automated cleanup for temporary/cached data.\par\vspace{-2pt}
    \begin{minipage}[t]{\linewidth}\vspace{0pt}
    \begin{lstlisting}[style=mystyle,language=Java,numbers=none,xleftmargin=0pt,framexleftmargin=0pt,framesep=2pt,aboveskip=0pt,belowskip=0pt]
// Cached data automatically expires.
void cacheData(String key, Object data, long ttl) {
  CacheEntry entry = new CacheEntry(data, System.currentTimeMillis() + ttl);
  cache.put(key, entry);}
// Periodically clean up expired cache.
void cleanupExpiredCache() {
  long now = System.currentTimeMillis();
  cache.entrySet().removeIf(entry -> entry.getValue().getExpiryTime() < now);}\end{lstlisting}\vspace{1pt}
    \end{minipage} &
    Retain data indefinitely or omit cleanup/expiry.\par\vspace{-2pt}
    \begin{minipage}[t]{\linewidth}\vspace{0pt}
    \begin{lstlisting}[style=mystyle,language=Java,numbers=none,xleftmargin=0pt,framexleftmargin=0pt,framesep=2pt,aboveskip=0pt,belowskip=0pt]
// Cached data retained permanently.
void cacheData(String key, Object data) {
  cache.put(key, data); // permanently retained
}
    

    
    \end{lstlisting}\vspace{1pt}
    \end{minipage} \\
    \hline
    \end{tabular}
\vspace{-8pt}
\end{table*}

\subsection{Taxonomy of Data Minimization Scenarios and Their Coding Practices}

Table~\ref{tab_dm_stage_scenarios} summarizes the identified scenarios across five data-handling stages in Android development, each scenario is accompanied by a representative \emph{DM-aligned} and \emph{DM-risky} coding practice.


\textbf{S1 - Permission Control.}
Permissions determine which system resources and user data that an application can access.
From a data minimization perspective, permission control delineates what data is authorized to access and when that access is requested.
\begin{itemize} [leftmargin = *]
    \item [1.1] \textbf{Permission Declaration Minimization (PDM)} suggests that applications declare in \emph{AndroidManifest.xml} only the permissions that are actually exercised in the code.
    Over-declaring permissions expands an app's authorized access scope and can increase privacy exposure if those additional capabilities are later exercised intentionally or inadvertently.
    This over-declaring has also been frequently reported in previous studies~\cite{Felt2011AndroidPD,Au2012PScoutAT,Backes2016OnDT}.

    \item [1.2] \textbf{Permission Request Minimization (PRM)} suggests that applications request permissions only when they are necessary for a specific feature at runtime. 
    Requesting permissions prematurely expands an app's authorized access scope beyond what the current feature requires, increasing unnecessary data access and the risk of misuse, and can also undermine user trust.
\end{itemize}


\textbf{S2 - Acquisition.}
When acquiring data (e.g., retrieving data from databases or file systems), the less data acquired, the lower the risk of misuse.

\begin{itemize} [leftmargin = *]
    \item [2.1] \textbf{Data Acquisition Minimization (DAM)} suggests that applications acquire only the minimum data necessary for functionality, selectively retrieving fields instead of all attributes or fields.
    Overly broad acquisition increases privacy exposure and can propagate to downstream processing, storage, or transmission even when only a subset of the retrieved fields is ultimately used.
\end{itemize}

\textbf{S3 - Transmission.}
In the data transmission stage, applications transform acquired data to implement functionality and may invoke third-party SDKs or services (e.g., a weather API).
From a data minimization perspective, once data crosses the app boundary, developers typically cannot control how external parties handle it.

\begin{itemize} [leftmargin = *]
    \item [3.1] \textbf{Data Transmission Minimization (DTM)} suggests that applications transmit only the minimum sensitive data necessary to external endpoints.
    For example, when saving a bookmark, an input URL may contain an optional email parameter that is not required for bookmarking.
    Removing such identifiers preserves functionality while reducing unnecessary data transmission.
\end{itemize}

\textbf{S4 - Storage.}
After data is processed, applications may need to store it for later use (e.g., sessions, preferences, or cached results).
From a data minimization perspective, storage decisions are critical because persisted data can be accessed long after the original feature execution and may be exposed through logs, backups, misconfigurations, or device compromise.
In this stage, we observe five recurring DM-critical development scenarios.

\begin{itemize} [leftmargin = *]
    \item [4.1] \textbf{Data Logging Minimization (DLM)} suggests that applications avoid recording sensitive information in logs.
    Failing to do so can leak PIIs (e.g., identifiers, tokens, or URLs) with embedded variables via log files, diagnostic exports, or third-party crash-reporting and analytics pipelines.

    \item [4.2] \textbf{Data Backup Minimization (DBM)} suggests that developers restrict the backup scope by explicitly constraining it through configuration files.
    Leaving backups broadly enabled could replicate sensitive or transient state into additional storage domains, increasing exposure and persistence beyond feature needs.


    \item [4.3] \textbf{Key Storage Minimization (KSM)} suggests that sensitive secrets (e.g., encryption keys and authentication keys) be stored in platform-protected storage such as Android KeyStore~\cite{android_keystore}.
    Persisting exportable key material (e.g., AES key) in local preferences or files makes it readable on compromised devices, which can enable unauthorized decryption or misuse of credentials.

    \item [4.4] \textbf{Biometric Handling Minimization (BHM)} suggests that applications use platform biometric APIs (e.g., \emph{BiometricPrompt}) that encapsulate biometric data within the system and expose only authentication outcomes.
    Directly generating, storing, or managing biometric templates in application code increases risk of irreversible exposure if the device or storage is compromised.

    \item [4.5] \textbf{Data Identifiability Minimization (DIM)} suggests that applications encrypt sensitive data at rest so that stored records cannot be directly read if the storage is exposed.
    If stored identifiers and credentials remain in plaintext, unauthorized access (e.g., through device compromise or unintended backup exports) can enable immediate account takeover.
\end{itemize}

\textbf{S5 - Retention.}
The data retention stage concerns how long stored data remains available.
From a data minimization perspective, the shorter the data persists on a device, the smaller the window for unauthorized access or unintended exposure.

\begin{itemize} [leftmargin = *]
    \item [5.1]\textbf{Retention Time Minimization (RTM)} requires developers to retain stored data only for the shortest time necessary for functionality, enforced through explicit expiry policies and automated cleanup mechanisms.
    For example, applying a TTL-based expiration policy to cached data ensures that it is automatically deleted after a defined period.
\end{itemize}


\section{Approach and Validation} \label{sec_approach}

Our formative study identifies ten DM-critical scenarios across five data-handling stages through a qualitative analysis of open-source Android applications.
Building upon these findings, we conduct a heuristic-based large-scale empirical analysis on a broader set of Android applications, aiming to obtain more comprehensive and generalizable insights.



\subsection{Approach Design}
Our approach captures DM coding practices through a unified set of heuristic-based indicators, each grounded in observable implementation evidence corresponding to DM-relevant scenarios.

We identify three categories of implementation evidence:
API usage, declarative configurations, and data flows.
First, many DM-relevant behaviors are reflected in API invocations, including those for data acquisition, encryption, key storage, biometric authentication, and data retention.
Second, some behaviors are specified in configuration artifacts (e.g., \emph{AndroidManifest.xml}, backup rules), which define how data is accessed, stored, and persisted, and may reveal over-declaration or overly broad scopes.
Third, certain behaviors arise from how data propagates through the program, such as flows from sensitive sources to sinks that expose or persist data (e.g., network transmission, logging).

To extract these evidence types, we apply corresponding static analysis techniques. 
For API usage, we construct call graphs to identify invoked APIs and their contexts, and apply pattern-based rules derived from our formative study (e.g., curated API lists and keyword matching). 
This enables detection of behaviors related to data acquisition (DAM), secure key management (KSM), biometric authentication (BHM), encryption (DIM), and data retention (RTM).

For \textbf{configuration-based evidence}, we analyze declarative artifacts to identify mismatches between declared and required behaviors. 
We derive rule-based checks from our formative study to detect issues related to permission declaration (PDM), permission request timing (PRM), and backup configurations (DBM). 
For example, for Permission Declaration Minimization (PDM), we extract declared permissions ($P_{\text{decl}}$) from the manifest and approximate required permissions ($P_{\text{req}}$) from invoked APIs.
If $P_{\text{decl}} \setminus P_{\text{req}} \neq \emptyset$, we report over-claimed permissions.
Similarly, for Permission Request Minimization (PRM), we analyze the control context of permission requests to identify premature requests.
For Data Backup Minimization (DBM), we inspect backup configurations to determine whether data inclusion is overly permissive.

For \textbf{data-flow-based evidence}, we perform static taint analysis, treating sensitive APIs as sources and predefined patterns as sinks (e.g., network transmission or logging).
This enables detection of behaviors related to data transmission (DTM) and data logging (DLM), where sensitive data is propagated to external or persistent endpoints. 
For example, the data transmission (DTM) indicator detects flows from sensitive sources to network sinks, while the data logging (DLM) indicator captures flows to logging APIs.

We implement call graph construction and taint analysis using FlowDroid~\cite{Arzt2014FlowDroidPC}, which provides context-, flow-, and lifecycle-aware analysis for Android applications. 
Additional details are available in our repository~\cite{our_repo}.

\begin{table}[t]
\centering
\caption{Effectiveness of our manual validation.}
\label{tab_effectiveness_validation}
\vspace{-10pt}
\scriptsize
\setlength{\tabcolsep}{4pt}
\resizebox{0.48\textwidth}{!}{%
\setlength{\aboverulesep}{0pt}
\setlength{\belowrulesep}{0pt}
\begin{tabular}{crrrr}
\toprule
\textbf{Indicator} & \textbf{\#APKs} & \textbf{\#Instances} & \textbf{Prec.\ (APK)} & \textbf{Prec.\ (Instances)} \\
\midrule
\rowcolor{gray!8}
PDM & 30 & 41 & 76.67\% & 70.73\% \\
\rowcolor{gray!18}
PRM & 30 & 34 & 83.33\% & 76.47\% \\
\rowcolor{gray!8}
DAM & 30 & 41 & 83.33\% & 87.80\% \\
\rowcolor{gray!18}
DTM & 30 & 90 & 90.00\% & 90.00\% \\
\rowcolor{gray!8}
DLM & 30 & 139 & 93.33\% & 93.53\% \\
\rowcolor{gray!18}
DBM & 30 & 30 & 100.00\% & 100.00\% \\
\rowcolor{gray!8}
KSM & 30 & 49 & 100\% & 100\% \\
\rowcolor{gray!18}
BHM & 30 & 61 & 100\% & 100\% \\
\rowcolor{gray!8}
DIM  & 30 & 63 & 93.33\% & 82.54\% \\ 
\rowcolor{gray!18}
RTM & 30 & 49 & 93.33\% & 87.23\% \\
\bottomrule
\end{tabular}%
}
\vspace{-15pt}
\end{table}

\subsection{Effectiveness Validation}

\subsubsection{Dataset and Execution}
We use AndroZoo~\cite{Allix2016AndroZooCM} as our data source to collect real-world Android applications for large-scale empirical analysis.
AndroZoo is a repository that provides metadata and APK files for millions of applications collected from various sources. 
As of December 28, 2025, it contains metadata for 9,154,054 applications and 22,608,573 APKs across multiple versions.
To focus on popular applications, we filter apps by maximum download count and retain those with at least 10 million downloads, resulting in 14,237 applications (989,782 APKs).
For each application, we select the latest APK, yielding a final dataset of 14,237 APKs.

We analyze all APKs on a Linux server with an Intel Core i9-9920X CPU @ 3.50,GHz and 188,GB RAM, using a per-APK timeout of 600,s.
The analysis completes in 58 days, with an average of 347.83\,s per APK among completed runs.
Overall, we successfully analyze 9,875 APKs (69.36\%).
Most failures are due to timeouts on large applications.
The remaining failures primarily result from malformed or incomplete APKs (e.g., missing or unparsable artifacts) and occasional pipeline crashes.

\subsubsection{Effectiveness}
To assess whether our indicators reflect genuine DM-relevant behavior, we estimate \emph{precision} via manual validation.
For each indicator, we randomly sample up to 30 flagged APKs.
Two authors independently decompile each APK using Jadx~\cite{Jadx} and inspect the relevant code to determine correctness. 
Disagreements are resolved through discussion.
Precision is computed as the fraction of sampled positives judged correct.
We do not estimate recall, as identifying false negatives would require exhaustive labeling of unflagged apps or code paths, which is infeasible at this scale. 
This focus on precision aligns with prior large-scale privacy analyses where full ground truth is prohibitively expensive~\cite{alecci2025toward}.

As shown in Table~\ref{tab_effectiveness_validation}, our indicators achieve high precision across most scenarios. 
Indicators based on explicit API usage or data flows (e.g., DTM, DLM, KSM, and BHM) consistently exceed 90\%, with several reaching 100\%. 
Permission-related indicators (PDM and PRM) exhibit comparatively lower precision (70–83\%), primarily due to incomplete API-to-permission mappings reused in our approach~\cite{Au2012PScoutAT,APERmapping,Backes2016OnDT}. 
This issue is further exacerbated by the use of recent APK versions, for which existing mappings are often outdated and fail to cover newly introduced or modified APIs. 
Constructing comprehensive API-to-permission mappings remains a well-known challenge, and no fully complete mapping currently exists~\cite{Felt2011AndroidPD,Backes2016OnDT}. 
As these mappings evolve, the precision of our approach is expected to improve accordingly. 
We next analyze the findings produced by these indicators to characterize data minimization practices in real-world Android applications.


\begin{figure}[t]
    \centering
    \includegraphics[width=0.85\columnwidth]{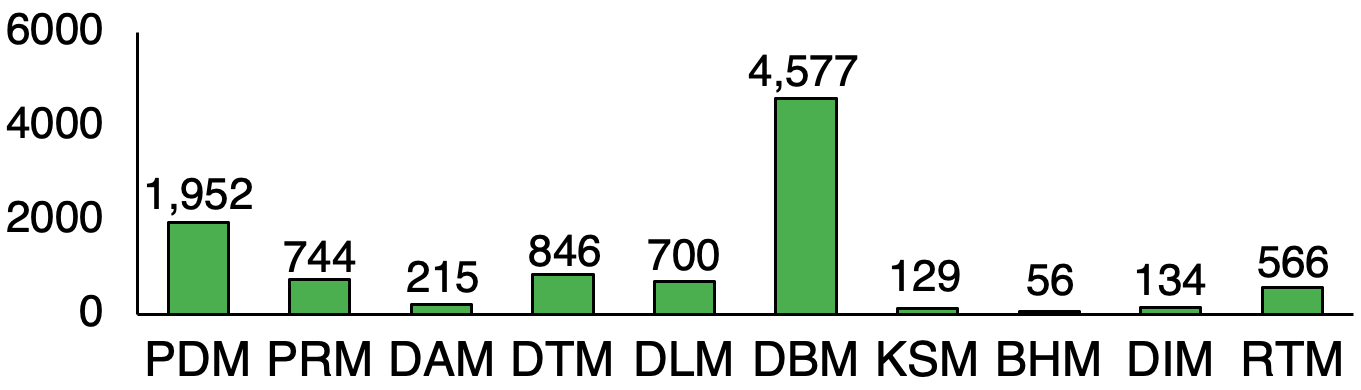}
    \vspace{-10pt}
    \caption{Per-indicator breakdown of data minimization practices across 9,875 analyzed APKs.}
    \label{fig_dm1_pre_indicator}
    \vspace{-15pt}
\end{figure}

\section{Results} \label{sec_results}

Our large-scale analysis identifies 7,548 APKs (76.44\% of 9,875) that exhibit data minimization–related evidence. 
Figure~\ref{fig_dm1_pre_indicator} shows the distribution of APKs across ten identified data minimization scenarios. 
In this section, we systematically examine how real-world applications align with or deviate from data minimization principles. 
Based on these findings, we distill a set of actionable coding practices and developer guidelines.



\subsection{Permission Control}

\begin{figure}[t]
    \centering
    \includegraphics[width=0.99\columnwidth]{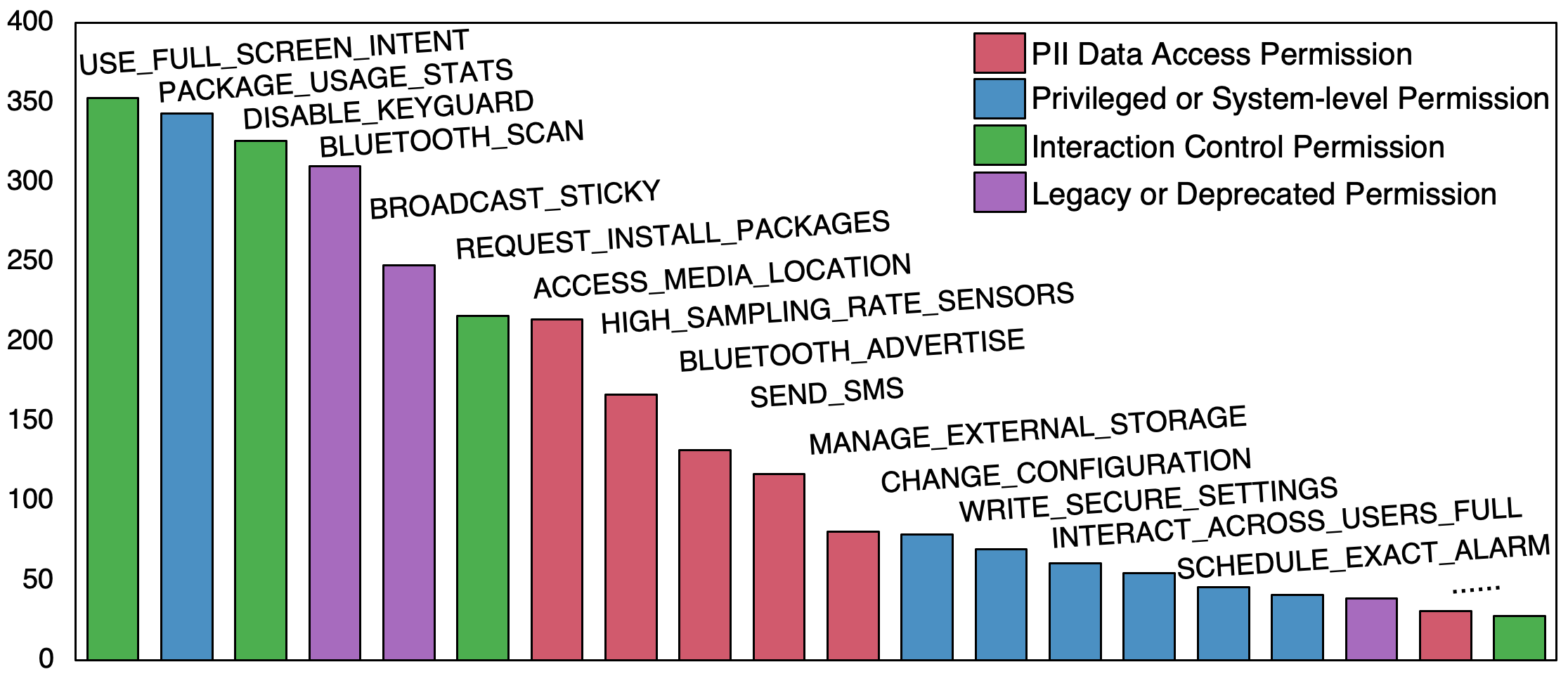}
    \vspace{-10pt}
    \caption{Occurrences of over-claimed permissions.}
    \label{fig_dm1_overclaimed_permissions_frequency}
     \vspace{-15pt}
\end{figure}

\noindent\underline{\textbf{Permission Declaration Minimization (PDM)}}.
We identify 1,952 apps (19.77\%) that declare permissions unused in the code, covering 113 distinct permission types and 3,497 over-claimed permission instances.
Figure~\ref{fig_dm1_overclaimed_permissions_frequency} shows a head-heavy distribution with a long tail.
A relatively small subset of permissions dominates the findings: 15 out of 113 permission types appear more than 50 times (13.27\%), while 77 types appear at most 10 times (68.14\%).
We group over-claimed permissions into four semantic categories.

The first category is \textit{PII data access permissions}, which expose sensitive user information such as identifiers and device signals (e.g., messaging, media location, and broad file storage access). 
This category accounts for 1,401 of 3,497 over-claimed instances (40.06\%) across 16 permission types. 
The second category is \textit{privileged or system-level permissions}, associated with system or signature-level capabilities (e.g., modifying system settings, controlling telephony state, or cross-user interaction). 
However, most apps are neither eligible nor required to request these permissions due to platform-enforced restrictions~\cite{AndroidPermission}. 
This suggests a lack of clear guidance on proper permission declaration. 
The third category is \textit{interaction control permissions}, which enable user-facing interruption or cross-app control (e.g., \emph{USE\_FULL\_SCREEN\_INTENT}). 
This category contributes 700 instances (20.02\%) across 14 permission types. 
Such over-declaration expands the application’s behavioral attack surface, enabling intrusive UI flows or unintended cross-app interactions. 
The last category is \textit{legacy or deprecated permissions}, capturing declarations retained due to historical carry-over. 
A representative example is \emph{BROADCAST\_STICKY}, which has known security risks and was deprecated in Android 5.0~\cite{StickyBroadcasts}.
However, it is still declared in 248 flagged APKs despite no observable usage in reachable code paths. 
This category contributes 295 instances (8.44\%) across 3 permission types.

\noindent\underline{\textbf{Permission Request Minimization (PRM)}}.
We identify 744 apps (7.56\%) with potentially premature permission requests across 32 permission types. 
The most frequently affected categories include notification access, camera usage, external storage, and fine-grained location. 
These applications request permissions during app initialization or early lifecycle callbacks, before any feature-level user interaction,
suggesting that requests are centralized in initialization logic or issued proactively for anticipated functionality (e.g., notifications, location services, or media access).
However, requesting permissions too early may negatively affect user experience and weaken user expectations regarding how permissions are used.



\find{\textbf{Summary of Coding Guidelines about Permission Request:}
\begin{itemize}[leftmargin=*]
\item Carefully reuse or merge manifest templates.
\item When removing features based on specific permissions, ensure the corresponding uses-permission entries are also removed.
\item Provide justification and reachability or traceability evidence for each declared permission before release. Use static analysis tools where possible.
\item Declare permissions only when required by the current functionality, rather than preemptively declaring all possible permissions at the initialization/launching stage.
\end{itemize}
}


\subsection{Data Acquisition}
\noindent\underline{\textbf{Data Acquisition Minimization (DAM)}}.
We identify 215 apps (2.18\%) that invoke at least one bulk-retrieval API without specific scoping.
Across these apps, we observe 246 invocations of 49 APIs that may retrieve all attributes instead of necessary data. 
We group these into three recurring patterns.
The most common scenario arises in app-local databases (SQLite/Room), where apps retrieve local data via SQL queries constructed through \emph{SQLiteQueryBuilder} or Room DAOs.
When queries are issued without column or row restrictions (e.g., \emph{SELECT *}), unnecessary data is retrieved.
Another scenario involves file and document metadata accessed via \emph{DocumentsProvider}. 
In this setting, queries materialize metadata into result cursors, and when no projection (i.e., the requested column set) is specified, providers may return all available metadata fields, including those not required by the application logic.
A DM-aligned approach is to enforce a restricted default projection, ensuring that only relevant fields are returned even when the caller provides \emph{null}.
For example, as shown in Listing~\ref{lst:penup_document_provider}, \emph{PENUP}~\cite{PENUP} defines a six-column default projection to limit exposed metadata.
\begin{lstlisting}[
  style=mystyle,
  language=Java,
  numbers=none,
  xleftmargin=0pt,
  framexleftmargin=0pt,
  basicstyle=\ttfamily\fontsize{6.5}{8}\selectfont,
  caption={Restricting the default projection in a DocumentsProvider to limit exposed metadata.},
  label=lst:penup_document_provider
]
public class BnRDocumentProvider extends DocumentsProvider {
// Define a restricted default projection to limit exposed metadata fields
  String[] defaultProjection = {
    "document_id", "mime_type", "_display_name",
    "last_modified", "flags", "_size"};
  @Override
  public final Cursor queryDocument(String id, String[] proj) {
// If caller does not specify a projection (i.e., requests all fields), fallback to the restricted default to enforce data minimization
    if (proj == null) { proj = defaultProjection; }
    MatrixCursor cursor = new MatrixCursor(proj);...}}
\end{lstlisting}
The third scenario occurs when apps query system-managed \emph{ContentProvider}s (e.g., \emph{MediaStore}, \emph{ContactsContract}) via \emph{ContentResolver.query()}.
If the parameter, the \emph{projection} (column filters) or \emph{selection} (row filters) is \emph{null}, it would retrieve all columns or rows, acquiring a large amount of unnecessary data.

\find{\textbf{Summary of Coding Guidelines about Data Acquisition:}
\begin{itemize}[leftmargin=*]
\item For database queries, explicitly specify required columns and apply row-level predicates.
\item When using SQLiteQueryBuilder, enforce column whitelisting via setProjectionMap() and enable setStrict(true).
\item For DocumentsProvider, define minimal default projections to ensure that only relevant metadata fields are returned when no projection is specified.
\item For system ContentProvider, always specify explicit projections, apply selection clauses to restrict rows, and use \emph{LIMIT} or pagination to avoid unnecessary result retrieval.
\end{itemize}
}

\subsection{Data Transmission}
\noindent\underline{\textbf{Data Transmission Minimization (DTM)}}.
We identify 846 APKs (8.57\%) with at least one sensitive data transmission flow, totaling 2,316 flows, where data originating from sensitive sources is transmitted to external network endpoints. 
These transmissions are initiated either by application code (first-party) or embedded third-party SDKs.
We characterize recurring transmission patterns based on data sources, destinations, and ownership, and analyze how these flows violate data minimization by transmitting data beyond what is strictly required for functionality.

For first-party data transmission, developers implement data collection and transmission logic within application code.
They might accidentally transmit data more than sufficient or without protection. 
For example, the social app \emph{Chatous}~\cite{Chatous} reads user preferences from \emph{SharedPreferences}, the Android ID via \emph{Settings.Secure}, and the system locale, and transmits them together in a single HTTP POST request. 
While the locale may support localization, including a persistent identifier such as Android ID introduces unnecessary tracking capability and could be replaced with a session-scoped token. 
Such flows may also involve transmitting user-generated content or account credentials without adequate protection.


Third-party data transmission is normally initiated by embedded third-party SDKs that read local data and transmit it to their own backends without explicit developer-defined transmission logic.
SDKs may invoke \emph{SharedPreferences.get*()} to read tokens, timestamps, and configuration data stored by the application, and forward them to advertising or analytics services. 
For example, the Chartboost SDK reads IAB Global Privacy Platform consent strings (\emph{IABGPP\_HDR\_GppString}) from \emph{SharedPreferences} and forwards them into its ad-tracking pipeline.
Beyond local storage, SDKs also collect device identifiers, system context information, and lists of installed applications. 
Although each of these data types may appear individually non-sensitive, their aggregation enables the construction of composite device fingerprints.
Furthermore, we observe cross-SDK data flows, wherein data accessed by one SDK is subsequently transmitted to another SDK’s processing pipeline. 
Such a phenomenon suggests that co-embedded SDKs may implicitly share data without explicit developer awareness or control.


\find{\textbf{Summary of Coding Guidelines for Data Transmission:}
\begin{itemize}[leftmargin=*]

\item Avoid transmitting device-stable identifiers (e.g., \emph{Settings.Secure.ANDROID\_ID}, IMEI, MAC) unless required, using resettable or session-scoped identifiers instead.

\item Construct network requests (e.g., \emph{OkHttp}, \emph{HttpURLConnection}) with only required fields, avoiding bundling unrelated data such as preferences, device information, or identifiers.

\item Avoid directly propagating data from sensitive sources (e.g., \emph{ContentResolver}, \emph{AccountManager}) to network sinks, applying filtering or transformation where possible.

\item Avoid storing sensitive data in shared storage (e.g., \emph{SharedPreferences}) accessible to third-party SDKs unless required.

\item Avoid reusing data collected for one feature (e.g., settings, consent flags) in network requests serving unrelated purposes such as advertising or analytics.

\end{itemize}
}

\subsection{Data Storage}

\noindent\underline{\textbf{Data Logging Minimization (DLM)}}.
We identify 3,010 sensitive-data-to-log flows across 700 APKs (7.09\%).
The logged data spans several categories, primarily including local file paths, network response content, location data, user input, and device identifiers.

A common pattern is logging \textit{internal file paths and directory information}, such as return values of \emph{Context.getFilesDir()} and \emph{Environment.getExternalStorageDirectory()}.
Although these logs are often used for debugging I/O operations, they expose the app’s internal storage structure and may reveal user-identifiable directory names (e.g., ``User/Alice/Document/...''.
Another frequent pattern is logging \textit{network response content}.
SDKs and app code read raw HTTP responses and write them to logcat for diagnostics.
Because responses may contain tokens, session identifiers, or personalized content, such logging can expose sensitive data that was not intended to be persisted locally.
We also observe logging of \textit{location data} and user input.
Apps and SDKs log precise geographic information and raw user input (e.g., via \emph{EditText.getText()}), which may include PII such as credentials, messages, or search queries.
Among all categories, logging user input poses a particularly high risk due to its direct exposure of sensitive information.
Finally, although less frequent, logging \textit{device identifiers} (e.g., IMEI or Advertising ID) carries disproportionate risk, as such identifiers are persistent and cannot be revoked once exposed.

\begin{figure}[t]
\centering
\begin{subfigure}[t]{0.5\columnwidth}
\centering
\includegraphics[width=\textwidth]{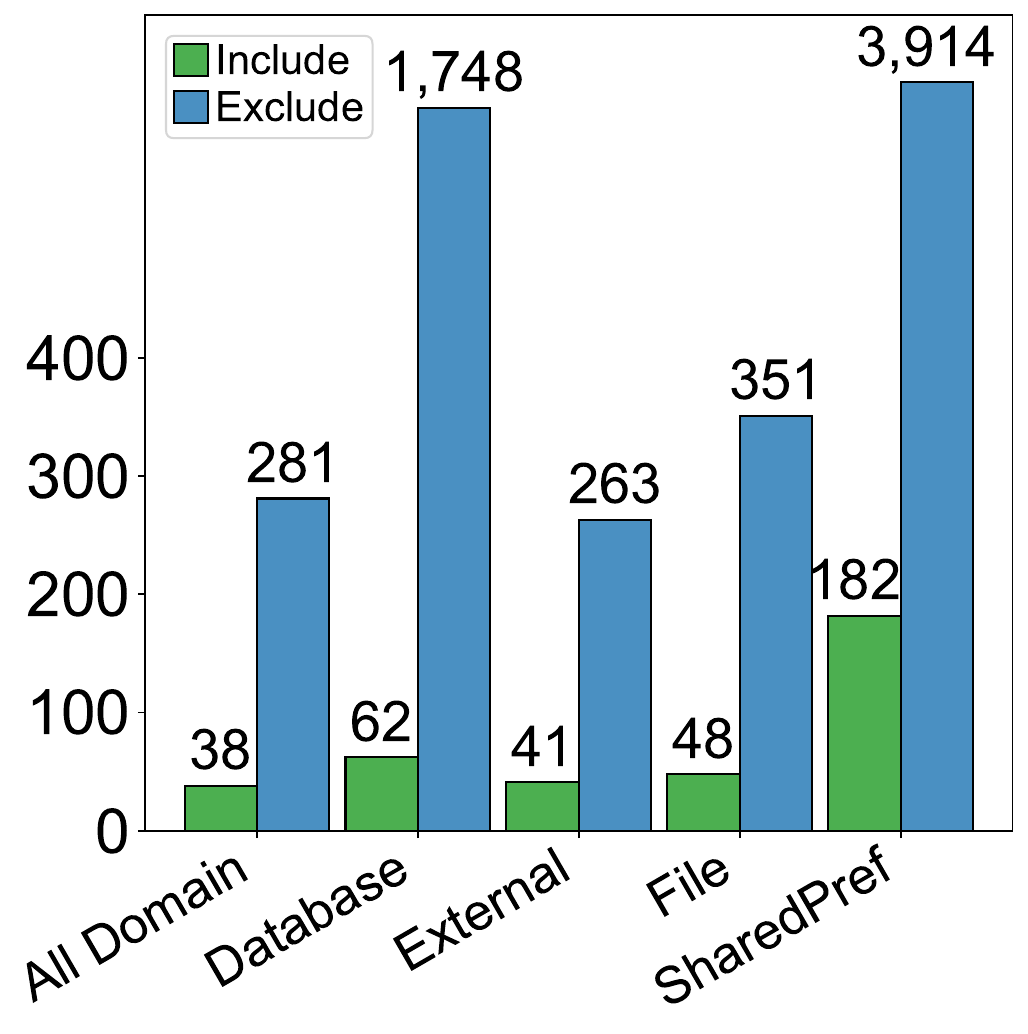}
\caption{\emph{cloud-backup}}
\label{fig_backup_cloud}
\end{subfigure}\hfill
\begin{subfigure}[t]{0.5\columnwidth}
\centering
\includegraphics[width=\textwidth]{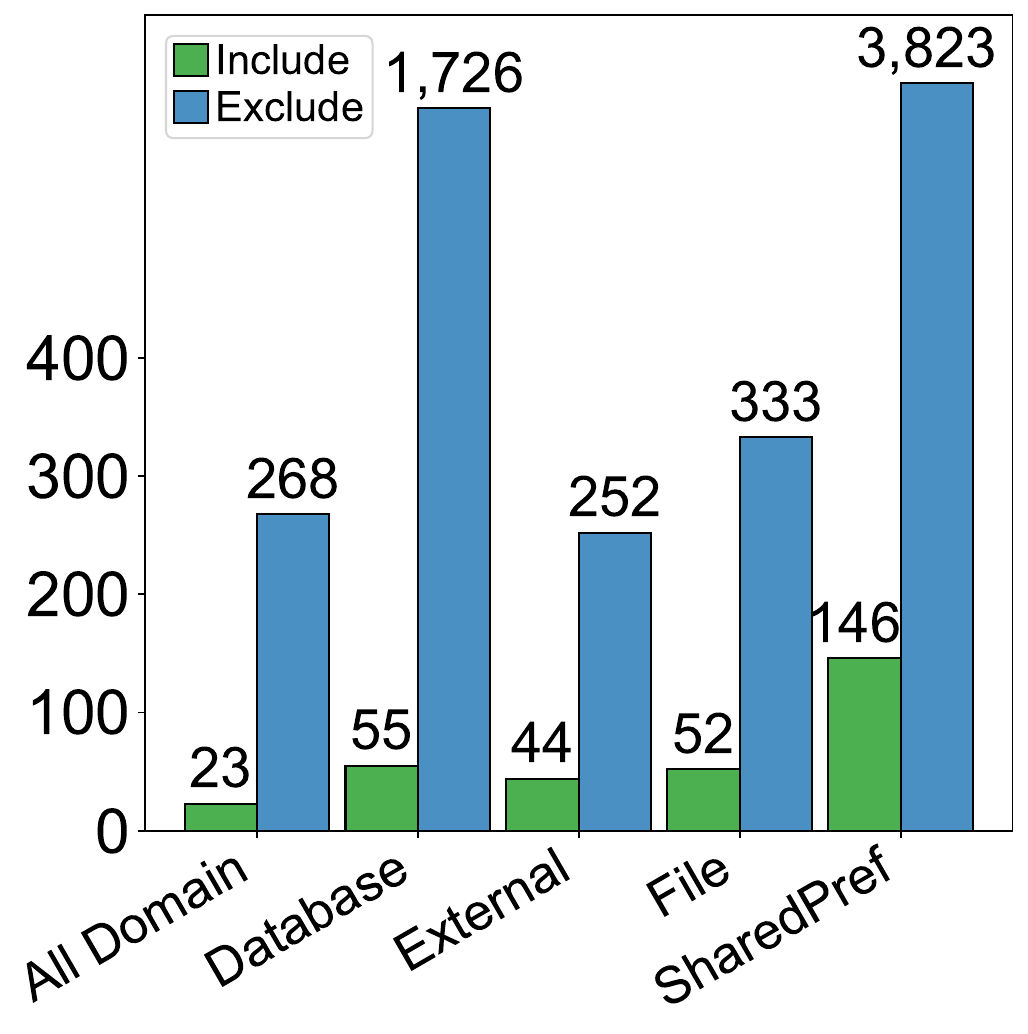}
\caption{\emph{device-transfer}}
\label{fig_backup_device}
\end{subfigure}
\caption{Include/exclude rules across backup domain types.}
\label{fig_backup_domain_comparison}
\end{figure}

\noindent\underline{\textbf{Data Backup Minimization (DBM)}}.
Among the 9,875 analyzed APKs, 4,345 apps leave backup enabled by default, resulting in automatic backup of all eligible data. 
In addition, 232 apps explicitly enable backup but do not define rules to restrict its scope. 
Overall, 4,577 apps (46.35\%) enable backup without specifying any rules, allowing unrestricted data backup. 
Android documentation recommends that developers explicitly define backup rules to control what data is included in backups~\cite{Backup}.
However, only 1,749 apps both enable backup and specify such rules. 
This indicates that currently developers do not care about data backup minimization.

We further analyze these 1,749 apps to understand how backup rules are specified.
We observe that \emph{exclude} rules significantly outnumber \emph{include} rules across both backup scopes (Figure~\ref{fig_backup_domain_comparison}).
This indicates a dominant \emph{blocklist-first} pattern, where developers rely on default full-backup behavior and selectively exclude certain data, rather than explicitly specifying the minimal set of data to retain.

The excluded data is dominated by SDK-generated artifacts and transient runtime state.
In particular, a large proportion of exclusion rules target data in shared preferences and databases associated with advertising, attribution, and analytics SDKs (e.g., AppsFlyer, Vungle, Airbridge).
These artifacts typically contain tracking configurations, tokens, and interaction logs that are unnecessary for app restoration and pose privacy risks if preserved.
Exclusion rules also frequently target cache directories and temporary files, including SDK caches and error logs, which are typically reconstructible and need not persist across device transfers.

In contrast, inclusion rules focus on core application data that supports user continuity, such as configuration databases, user profiles, and application state stored in databases or shared preferences.
We also observe inclusion of user-generated or user-visible data, such as saved content and documents in external storage.

\find{\textbf{Summary of Coding Guidelines about Logging and Backup:}
\begin{itemize}[leftmargin=*]
\item Avoid logging raw user input, credentials, or device identifiers.
\item Remove or disable logging statements in release builds.
\item Define specific backup rules instead of relying on the default full-backup to avoid unnecessary data remnants.
\item Avoid store authentication credentials, secret tokens, or persistent identifiers in backups.
\end{itemize}
}

\noindent\underline{\textbf{Key Storage Minimization (KSM)}}.
We identify 227 KeyStore API usages across 129 APKs (1.31\%). 
The most common scenario is \textit{TLS certificate trust chain setup}, where apps load system trust anchors or custom CA certificates during network initialization (e.g., \emph{Application.onCreate()}). 
This pattern is DM-neutral, as it supports secure communication but does not directly protect application data. 
A second scenario involves \textit{hardware-backed key management via \emph{AndroidKeyStore}}.
In this pattern, apps generate or retrieve cryptographic keys within the \emph{AndroidKeyStore} provider, ensuring that key material is protected by hardware-backed isolation and cannot be directly accessed by application code.
This represents a DM-aligned practice, as it minimizes key exposure.
A third scenario is a \textit{file-backed keystore with hardcoded credentials}, which represents a DM-risky anti-pattern.
In this case, apps use a software-backed keystore and persist key material to local storage with a hardcoded password.
For example, the file management app \emph{WD2Go}~\cite{WD2Go} uses \emph{KeyStore.getDefaultType()} instead of \emph{AndroidKeyStore}:
\begin{lstlisting}[
  style=mystyle,
  language=Java,
  numbers=none,
  xleftmargin=0pt,
  framexleftmargin=0pt,
  basicstyle=\ttfamily\fontsize{6.5}{8}\selectfont,
  caption={File-backed keystore with hardcoded password, exposing key material to disk},
  label=lst:WD2Go
]
// Using default (file-based) keystore instead of "AndroidKeyStore"
KeyStore ks = KeyStore.getInstance(KeyStore.getDefaultType());
char[] pw = "hardcoded_password".toCharArray();
ks.load(null, pw);
SecretKey key = KeyGenerator.getInstance("AES").generateKey();
ks.setEntry(KEY_ALIAS,new SecretKeyEntry(key),new PasswordProtection(pw));
new FileOutputStream(file); // key saved to disk
\end{lstlisting}
Since both the key file and its password reside on disk, any process with file-system access can extract the key material, defeating the purpose of using a keystore and increasing the risk of compromise.

\find{\textbf{Summary of Coding Guidelines about Key Storage:}
\begin{itemize}[leftmargin=*]
\item Use hardware-backed keystores (e.g., \emph{AndroidKeyStore}) to prevent key materials from being exposed to application memory.
\item Do not store cryptographic keys or passwords in application-accessible storage, especially hard-coded credentials.
\item Remove or invalidate keys when they are no longer needed (e.g., on logout or credential reset).
\item Avoid reusing keys across multiple functionalities.
\end{itemize}
}

\noindent\underline{\textbf{Biometric Handling Minimization (BHM)}}.
We identify 113 biometric API invocations across 56 APKs, covering \emph{BiometricPrompt} (45), \emph{BiometricManager} (36), and deprecated \emph{FingerprintManagerCompat} (32). 
Biometric handling is consistently DM-aligned, as no app directly processes raw biometric data. 
Instead, all apps delegate acquisition and matching to the Android platform.
The most common scenario is \textit{biometric-gated authentication}, where apps use \emph{BiometricPrompt.authenticate()} (or legacy \emph{FingerprintManagerCompat}) to gate access to sensitive features such as login or payment confirmation.
For example, the banking app \emph{Canara ePassbook}~\cite{ePassbook} triggers biometric login as follows:
\begin{lstlisting}[
    style=mystyle,
    language=Java,
    numbers=none,
    xleftmargin=0pt,
    framexleftmargin=0pt,
    basicstyle=\ttfamily\fontsize{6.5}{8}\selectfont,
    caption={Biometric-gated login via BiometricPrompt (Canara ePassbook)},
    label=lst:biometric_canara
]
// Create a platform-managed biometric prompt for login
BiometricPrompt prompt = new BiometricPrompt(this, getMainExecutor(this), callback);
BiometricPrompt.PromptInfo info = new BiometricPrompt.PromptInfo.Builder()
        .setTitle("Touch ID for " + appName)
        .setDescription("Place your finger to login")
        .setNegativeButtonText("Cancel").build();
// Start biometric authentication for the login flow
prompt.authenticate(info);
\end{lstlisting}
In this pattern, the app never touches fingerprint minutiae or facial feature vectors.
It only receives a success or failure callback from the platform.
This system-mediated design inherently enforces data minimization at the biometric data level.

The second scenario is \textit{biometric-bound cryptographic operations}, where apps combine \emph{BiometricPrompt} with an \emph{AndroidKeyStore}-backed \emph{CryptoObject}. 
For example, \emph{PalmPay}~\cite{PalmPay} calls \emph{authenticate(promptInfo, cryptoObject)}, ensuring cryptographic operations are gated by biometric authentication. 
Similarly, some apps generate keys in \emph{AndroidKeyStore} and wrap them in a \emph{CryptoObject} before authentication, ensuring keys remain hardware-isolated and usable only after verification.
This pattern provides the strongest guarantee: cryptographic material remains hardware-isolated and is released only upon live biometric confirmation.

We also observe that 27 APKs (48.21\%) still use deprecated \emph{FingerprintManagerCompat}. 
While this does not introduce DM risk, migrating to \emph{BiometricPrompt} enables stronger guarantees, such as Class~3 enforcement and explicit \emph{CryptoObject} binding.

\find{\textbf{Summary of Coding Guidelines about Biometric Data:}
\begin{itemize}[leftmargin=*]
\item Use \emph{BiometricPrompt} instead of the deprecated \emph{FingerprintManagerCompat} to benefit from unified multi-modal support and Class~3 biometric enforcement.
\item Bind sensitive cryptographic operations to biometric authentication via \emph{CryptoObject} to ensure keys are only usable after live user verification.
\item Never attempt to access, store, or transmit raw biometric data; rely entirely on the platform-mediated authentication callback.
\item Cancel ongoing biometric sessions in lifecycle callbacks (e.g., \emph{onPause()}) to prevent stale authentication state.
\end{itemize}
}

\noindent\underline{\textbf{Data Identifiability Minimization (DIM)}}.
We study encryption practices that reduce sensitive data exposure and limit identifiability risks. 
Across 134 APKs, we identify 318 encryption-related API invocations, the majority of which originate from application code rather than third-party SDKs. 
These observations reveal two recurring scenarios based on encryption targets.

The dominant scenario is \textit{transit encryption of network payloads}, where apps encrypt request bodies, telemetry, and serialized objects before transmission. 
The most common implementation pattern is an app-wrapped encryption method that delegates to \emph{javax.crypto.Cipher} with AES (34 occurrences) or RSA (8).
For example, the quiz app \emph{Qureka} implements a full AES pipeline with \emph{SecretKeySpec} and \emph{Cipher.getInstance("AES")} to encrypt WebSocket payloads before sending.
The second scenario is \textit{at-rest encryption of local data}, covering device identifiers, tokens and sessions, credentials, and files or streams. 
A strong practice is platform-backed encryption using \emph{AndroidKeyStore}: the banking app \emph{Citibanamex}~\cite{Citibanamex} generates AES keys within \emph{AndroidKeyStore} and encrypts database earmarks using \emph{AES/CBC} with proper IV handling and key lifecycle management (key regeneration and deletion).
However, only 9 records reference \emph{AndroidKeyStore} for key storage, and only one uses AES-GCM (authenticated encryption), suggesting that hardware-backed key protection remains rare even among apps that do encrypt.
We also observe weak or deprecated cryptographic schemes, including ECB mode (7 cases) and DES (5). 
For instance, the e-commerce app \emph{LightInTheBox} uses \emph{DES/CBC} with a key derived from a hardcoded secret, while another app uses \emph{DES/ECB} with a hardcoded 8-byte key:
\begin{lstlisting}[
    style=mystyle,
    language=Java,
    numbers=none,
    xleftmargin=0pt,
    framexleftmargin=0pt,
    basicstyle=\ttfamily\fontsize{6.5}{8}\selectfont,
    caption={Weak encryption: DES/ECB with a hardcoded key},
    label=lst:des_ecb
]
SecretKeySpec key = new SecretKeySpec(
    "??x?D@?w".getBytes(), "DES");    // hardcoded 8-byte key
Cipher cipher = Cipher.getInstance(
    "DES/ECB/PKCS7Padding", "BC");    // deprecated algorithm + ECB
cipher.init(Cipher.DECRYPT_MODE, key);
return new String(cipher.doFinal(ciphertext));
\end{lstlisting}
Both DES (56-bit effective key length) and ECB mode (no diffusion across blocks) are considered cryptographically broken, and their use negates the DM benefit that encryption is intended to provide.

\find{\textbf{Summary of Coding Guidelines about Data Identifiability:}
\begin{itemize}[leftmargin=*]
\item Use authenticated encryption modes (e.g., AES-GCM) instead of unauthenticated modes (CBC, ECB) to ensure  integrity.
\item Store encryption keys in \emph{AndroidKeyStore} rather than in application-accessible storage or hard-coded constants.
\item Avoid deprecated algorithms (DES, 3DES) and insecure modes (ECB); Use AES-256 with proper IV/nonce management.
\end{itemize}
}

\subsection{Data Retention}
\noindent\underline{\textbf{Retention Time Minimization (RTM)}}.
We identify explicit data-cleanup mechanisms in 566 APKs (5.73\%), comprising 1,172 cleanup API invocations, making it the most widely adopted DM-related practice.
We summarize four recurring data cleanup practices.
The most common is \textit{cache eviction}, where apps invoke \emph{clearCache()}, \emph{evictAll()}, or \emph{trimToSize()} to remove in-memory, disk, or WebView caches. 
While this reduces the data footprint, it mainly targets reconstructible data (e.g., HTTP caches, thumbnails), limiting its direct DM impact.
The second is \textit{user history and session cleanup}, where apps delete browsing history, messages, or session state from local storage.
The third practice is \textit{time-based database record expiry}, where apps delete records that have exceeded a defined retention window.
For instance, \emph{Telegram}~\cite{Telegram} implements an expiration-based cleanup for fact-check records: \emph{DELETE FROM fact\_checks WHERE expires > currentTimeMillis()}.
Similarly, the restaurant reservation app \emph{OpenTable}~\cite{OpenTable} removes stale location preferences by individually clearing latitude, longitude, name, and timestamp fields from \emph{SharedPreferences}.
This approach represents the strongest DM practice, as it enforces an explicit retention limit on persistent data.
The fourth and rarest practice is \textit{account and credential reset}, where apps clear authentication tokens, session keys, or configuration state upon logout or credential change.
For example, the \emph{LimeIPTV} app removes stored hash sums and validity timestamps via \emph{preferences.edit().remove(key).apply()}.
Despite its significance, it is observed in only 2 analyzed records, indicating that credential cleanup remains an afterthought for most developers.

\find{\textbf{Coding Guidelines about Data Retention:}
\begin{itemize}[leftmargin=*]
\item Implement cleanup (e.g., \emph{clearCache()}) for all locally stored sensitive data, including database records, cached tokens, session state, user-generated content, browsing history, location data, and authentication credentials.
\item For events about logout, account deletion, and credential reset events, implement credential and session cleanup.
\item Adopt a systematic retention policy that audits all local storage locations (\emph{SharedPreferences}, SQLite databases, internal/external files) rather than relying on isolated point fixes.
\end{itemize}
}


\section{Data Minimization in LLM-based Code Generation} \label{sec_DM_in_LLMs}

Large language models (LLMs) have fundamentally transformed the paradigm of code generation in software development, enabling developers to produce code in an unprecedentedly autonomous and productive manner, widely known as ``vibe coding''.
LLMs and LLM-based code generation tools are trained on vast corpora of existing code, which is accumulated over decades through contributions from developers and open-source communities. 
As such, these models inherently learn statistical patterns from real-world implementations.
In our study, we observe both DM-aligned and DM-risky coding practices in real-world Android applications, with the latter being more prevalent. 
Given that LLMs are trained on such data, they are likely to capture not only best practices but also suboptimal or non-compliant patterns. 
Consequently, LLMs may not only reproduce but potentially amplify DM-risky coding practices during the LLM-based code generation.


Therefore, we investigate two key hypotheses: 1) whether LLMs reproduce DM-risky coding patterns in generated Android applications; and 2) whether incorporating the summarized coding guidelines derived from our empirical findings can effectively steer code generation toward DM-aligned implementations.

\paragraph{Experimental Setup.}
We manually construct 30 prompts to simulate LLM-based code generation in Android development. 
For each of the ten DM-related coding scenarios mentioned in Section~\ref{sec_results}, we manually craft three prompts that generally describe the main functionalities of the Android app, resulting in 30 prompts in total.
For example, for data acquisition minimization, one prompt asks the model to generate an Android app that shows nearby weather information based on the user's current location, which could involve location permission requests, data acquisition, and remote data transmission.
In the prompt, we ask the model to generate a complete Android application, including source code and necessary configuration files.
The detailed prompts are available in our code repository~\cite{our_repo}.
As for the models, we evaluate three representative LLMs, GPT-5.2~\cite{GPT52}, Claude-4.5-Sonnet~\cite{claude45sonnet}, and Gemini-2.5-Pro~\cite{gemini25pro}, along with the mainstream agentic coding tool, Cursor~\cite{Cursor}, to simulate the ``vibe coding'' paradigm.
After running the basic prompts, we further incorporate data minimization guidelines derived from Section~\ref{sec_results} into the prompts and re-run generation.
We then run the analysis approach (mentioned in Section~\ref{sec_approach}) and manually examine the identified data handling scenarios.

\begin{table}[t]
    \caption{Data minimization risky instances in LLM-generated Android applications. }
    \label{tab_ai_dm_violations}
    \centering
    \setlength{\aboverulesep}{0pt}
    \setlength{\belowrulesep}{0pt}
    \resizebox{\columnwidth}{!}{%
    \begin{tabular}{l *{10}{c} c} 
    \toprule
    \textbf{Generation Setting} & \rotatebox{90}{\textbf{PDM}} & \rotatebox{90}{\textbf{PRM}} & \rotatebox{90}{\textbf{DAM}} & \rotatebox{90}{\textbf{DTM}} & \rotatebox{90}{\textbf{DLM}} & \rotatebox{90}{\textbf{DBM}} & \rotatebox{90}{\textbf{KSM}} & \rotatebox{90}{\textbf{BHM}} & \rotatebox{90}{\textbf{DIM}} & \rotatebox{90}{\textbf{RTM}} & \rotatebox{90}{\textbf{Total}} \\
    \midrule
    \textit{Basic Prompt} & & & & & & & & & & & \\
    \/\/\/  Gemini-2.5-Pro       & \hmc{3} & \hmc{5} & \hmc{5} & \hmc{2} & \hmc{5} & \hmc{20} & \hmc{2} & \hmc{0} & \hmc{3} & \hmc{1} & \hmc{46} \\
     \/\/\/ GPT-5.2               & \hmc{3} & \hmc{4} & \hmc{3} & \hmc{1} & \hmc{4} & \hmc{28} & \hmc{2} & \hmc{0} & \hmc{5} & \hmc{8} & \hmc{57} \\
     \/\/\/ Claude-4.5-Sonnet    & \hmc{5}  & \hmc{5} & \hmc{4} & \hmc{1} & \hmc{4} & \hmc{28} & \hmc{3} & \hmc{0} & \hmc{6} & \hmc{3} & \hmc{59} \\
     \/\/\/  Cursor (Codex)       & \hmc{2} & \hmc{5} & \hmc{5} & \hmc{2} & \hmc{2} & \hmc{28} & \hmc{2} & \hmc{0} & \hmc{6} & \hmc{5} & \hmc{57} \\
    \midrule
    \textit{Basic Prompt + DM Guidelines} & & & & & & & & & & & \\
    \/\/\/ Gemini-2.5-Pro        & \hmc{0} & \hmc{0} & \hmc{0} & \hmc{0} & \hmc{0} & \hmc{0} & \hmc{0} & \hmc{0} & \hmc{0} & \hmc{0} & \hmc{0} \\
    \/\/\/ GPT-5.2               & \hmc{0} & \hmc{0} & \hmc{0} & \hmc{0} & \hmc{0} & \hmc{0} & \hmc{0} & \hmc{0} & \hmc{0} & \hmc{0} & \hmc{0} \\
    \/\/\/ Claude-4.5-Sonnet     & \hmc{0}  & \hmc{0} & \hmc{0} & \hmc{0} & \hmc{0} & \hmc{0} & \hmc{0} & \hmc{0} & \hmc{0} & \hmc{0} & \hmc{0} \\
    \/\/\/ Cursor (Codex)      & \hmc{0} & \hmc{0} & \hmc{0} & \hmc{0} & \hmc{0} & \hmc{0} & \hmc{0} & \hmc{0} & \hmc{0} & \hmc{0} & \hmc{0} \\
    \bottomrule
    \end{tabular}%
    }
\end{table}

\paragraph{Observations.}
Table~\ref{tab_ai_dm_violations} presents the number of DM-related coding practices observed in the generated applications under different settings. 
When generating code solely based on functional requirements in the basic prompts, all approaches produce a substantial number of DM-risky coding practices. 
Across the 30 scenarios, Gemini-2.5-Pro produces 46 instances of DM violations, GPT-5.2 produces 57 instances, Claude-4.5-Sonnet produces 59 instances, while Cursor produces  57 instances.
Among different categories, data backup minimization (DBM) violations occur most frequently, indicating that LLMs often overlook backup-related privacy configurations. 
In contrast, all approaches consistently exhibit correct practices in biometric handling minimization (BHM), suggesting that certain security-sensitive scenarios are better learned by LLMs.
Notably, this distribution closely aligns with our large-scale empirical findings in Section~\ref{sec_results}, where real-world applications also demonstrate strong compliance in BHM but frequently neglect backup-related configurations. 
This observation suggests that DM-risky coding practices prevalent in real-world applications are implicitly learned and reproduced by LLMs during code generation.

After incorporating DM-aware guidelines, the number of DM violations is reduced to zero across all evaluated approaches. 
These results indicate that the issue lies not in the capability of LLMs, but in the lack of explicit awareness of data minimization requirements during generation. 
With appropriate guidance, LLMs can consistently produce high-quality, privacy-compliant implementations.

\paragraph{Findings.} 

Results demonstrate that LLMs act as amplifiers of existing coding practices. 
DM-risky practices prevalent in real-world applications are implicitly learned by LLMs and subsequently propagated into LLM-based development workflows and LLM generated program. 
At the same time, LLMs also retain the capacity to reflect improved practices when appropriately guided.
Incorporating DM-aware coding guidelines proves highly effective in steering LLMs toward generating privacy-compliant implementations. 
Although LLMs inherit both strengths and weaknesses from their training corpus, their outputs can be systematically improved through principled, domain-specific guidance, thereby improving the coding quality in terms of better privacy-compliance.







\paragraph{Tension between Productivity and Regulatory Compliance}
Modern software development increasingly emphasizes rapid feature delivery and iterative prototyping.
Thus, developers often prioritize functional correctness, while non-functional requirements such as privacy compliance receive less attention. 
This imbalance gives rise to an inherent tension between productivity and regulatory compliance.
The emergence of LLM-based code generation simultaneously introduces both challenges and opportunities in this context.
On one hand, this tension is further amplified in LLM-driven workflows, particularly under the ``vibe coding'' paradigm.
When developers primarily evaluate generated code based on functional correctness, DM-risky patterns may be repeatedly introduced and propagated.
On the other hand, when properly configured and guided, LLMs have the potential to alleviate this tension. 
By incorporating explicit privacy-aware constraints and guidelines into the generation process, LLMs can assist developers in adhering to non-functional requirements without requiring substantial additional effort.

\section{Related Work}

Prior work in the Android ecosystem has extensively studied coding practices for robust application development, including permission handling~\cite{Felt2011AndroidPD, Au2012PScoutAT}, lifecycle-aware programming~\cite{sun2023taming, Arzt2014FlowDroidPC}, and correct API usage~\cite{McDonnell2013AnES, MannanUCodeSmell, sun2021taming, sun2023demystifying}. 
Static analysis identifies bug patterns and anti-patterns in apps, improving code quality~\cite{Arzt2014FlowDroidPC,Liao2025ACS,GuoResourceLeak}. 
While these efforts target functional correctness~\cite{sun2022mining}, performance~\cite{Liao2025ACS}, or security~\cite{Zhangpermission}, they largely overlook privacy-compliance coding practices.
To address data protection concerns, privacy engineering integrates privacy requirements into software development~\cite{Sangaroonsilp2025ASO,Bednar2018EngineeringPB,Hadar2017PrivacyBD,Maalej2014UsAT,Ferreira2023RuleKeeperGP}. 
Existing work includes privacy-by-design approaches, which remain high-level and lack implementation guidance~\cite{AndradePbD,Hadar2017PrivacyBD,Bednar2018EngineeringPB}, 
policy-compliance studies detecting inconsistencies between policies and behavior~\cite{Zimmeck2016AutomatedAO,Slavin2016TowardAF,Yu2018PPCheckerTA}, 
and formal or DSL-based approaches for automated compliance checking that require significant effort~\cite{Caramujo2018RSLIL4PrivacyAD,Young2011CommitmentAT,Colesky2016PrivacySB}.
Despite these advances, they offer limited support for implementing privacy requirements at code level, leaving a gap in actionable privacy-compliance coding practices.

\section{Conclusions}
\label{sec:conclusions}

In this work, we operationalize the high-level privacy principle of data minimization into concrete coding practices for Android development.
Through a formative study of 1,114 open-source applications, we identify ten recurring data minimization scenarios across five data-handling stages.
We then analyze 9,875 real-world APKs and distill 31 actionable coding guidelines for privacy-compliant development.
Our evaluation of LLM-based code generation shows that state-of-the-art models reproduce risky practices, inheriting and amplifying patterns from real-world code.
Incorporating our guidelines into generation eliminates these issues across all evaluated models.
Our work advocates addressing privacy at the source code level, enabling more reliable compliance in both human and AI-assisted programming.

\noindent \textbf{Data Availability Statement}. The dataset, source code and experimental results are available in our artifact repository:~\cite{our_repo}.

\bibliographystyle{ACM-Reference-Format}
\bibliography{references}



\end{document}